\documentclass[twocolumn,showpacs,preprintnumbers,prb,amsmath,amssymb]{revtex4}

\usepackage{graphicx}
\usepackage{dcolumn}
\usepackage{bm}

\begin{document}

\title{The magnetic form factor of iron in SrFe$_2$As$_2$}

\author{Y. Lee, David Vaknin, Haifeng Li, Wei Tian, Jerel L. Zarestky, N. Ni, S. L. Bud$'$ko, P. C. Canfield, R. J. McQueeney, and B. N. Harmon}

\affiliation{Ames Laboratory and Department of Physics and Astronomy, Iowa State University, Ames, Iowa 50011, USA}


\date{Submitted 25 November 2009; Accepted 19 January 2010 by PRB Rapid Communication}

\begin{abstract}
The iron magnetic form factor in SrFe2As2 has been determined by neutron diffraction and by density functional theory (DFT).  As noted previously, the magnitude of the calculated moment using DFT is sensitive to the Fe-As distance.  However, the shape of the calculated form factor is practically insensitive to the Fe-As distance, and further we show that the form factor closely resembles that of bcc iron, and agrees well with experiment. The spin density exhibits some anisotropy due to geometry and As hybridization.

\end{abstract}


\maketitle

The magnetism and nature of the iron-3d electronic states in FeAs based pnictide superconducting materials is of interest. The states near the Fermi level are involved in the superconductivity and are also responsible for the sensitivity of the magnetic moment to small structural changes. At first glance the proximity of the superconducting states to an antiferromagnetic (AF) ordered state suggests a similarity to cuprates; however the Mott-Hubbard AF insulator state in the undoped cuprates is not amenable to density 
functional theory (DFT) approaches\cite{num3,num4}, whereas DFT has been remarkably helpful in elucidating the electronic and magnetic behavior of FeAs related materials. As one example, we note that DFT calculations for the ambient and high pressure phase of CaFe$_2$As$_2$ predicted the disappearance of the magnetic moment in the collapsed phase.\cite{num5,num6} Mazin \textit{et al.} remark in their study of LaFeAsO$_{1-x}$F$_x$, ``magnetism in this compound 
is very itinerant, making the calculated magnetic energies and moments extremely sensitive to the approximations used 
and to fine details of the crystal structure''. They recommend the LDA functional and using the theoretical As 
positions for ``best results''.\cite{num7} In this work we investigate the spatial extent of spin density by both measuring the magnetic form factor and calculating it using the experimental and also the theoretical As positions.

\begin{figure}[bp]
	\centering
   \includegraphics[width=0.40\textwidth]{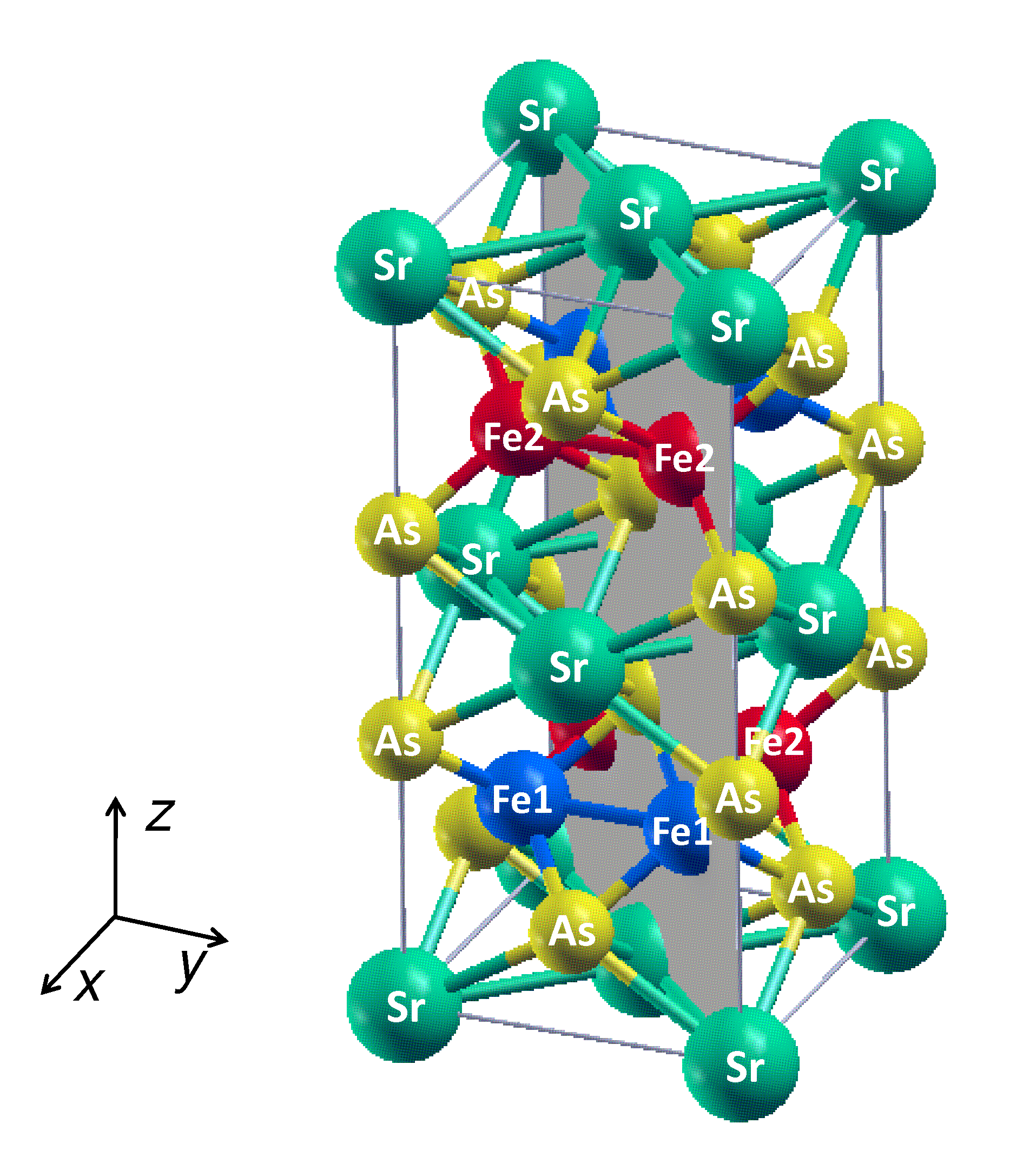}
	\caption{(Color online) SrFe$_2$As$_2$ orthorhombic crystal structure: the green spheres are Sr atoms, yellow are As atoms and both blue(Fe1) and red(Fe2) are Fe atoms but represent different magnetic moment directions. The gray plane is the one that was chosen for the spin density contour plot of Fig.\ \ref{spin2d1}.}
	\label{struct1}
\end{figure}

The orthorhombic crystal structure and magnetic ordering found in SrFe$_2$As$_2$ is illustrated Fig.\ \ref{struct1}. Above $\sim$200 K the 
crystal is tetragonal, but undergoes a simultaneous structural and AF transition below this temperature. The experimental 
value of the magnetic moment on each iron site in the AF state of SrFe$_2$As$_2$ has been estimated using neutron powder 
diffraction to be 1.01(3) $\mu_\texttt{B}$\cite{num10}, and 0.94(4) $\mu_\texttt{B}$\cite{num11}. There is some discussion if the AF/structure 
transition is first or second order\cite{num12, num13} but it is not relevant to our measurements taken at low temperatures ($\sim$10 K).

A high-quality SrFe$_2$As$_2$ single crystal synthesized by the FeAs flux growth technique \cite{Ni2008-1} was selected for this study. This crystal was also used in a recent neutron diffraction study of the nature of the magnetic and structural phase transitions in SrFe$_2$As$_2$.\cite{Li2009} The crystallinity and purity were characterized by Laue back-scattering, X-ray powder diffraction, magnetization and resistivity measurements. The mosaic is 0.29(1)$^\texttt{o}$ full width at half maxima for the (0 0 8) reflection.\cite{Li2009} The elastic neutron scattering measurements were carried out on the HB-1A fixed-incident-energy (14.6 meV) triple-axis spectrometer using a double pyrolytic graphite (PG) monochromator (located at the High Flux Isotope Reactor, HFIR, at the Oak Ridge National Laboratory, USA). Two highly oriented PG filters, one after each monochromator, were used to reduce the $\lambda$/2 contamination. The beam collimation throughout the experiment was kept at 48$'$-48$'$-sample-40$'$-136$'$. The single crystal ($\sim$ 15 mg) was wrapped in Al foil and sealed in a He-filled Al can which was then loaded on the cold tip of a closed cycle refrigerator with (\emph{H} 0 \emph{L}) in the \emph{Fmmm} symmetry as the scattering plane.

The magnetic form factor $F(\vec{K)}$ is a Fourier transformation of spin density at a given $\vec{K}$ vector and is given as
\begin{equation}
	F(\vec{K})=\int_{cell}S(\vec{r})e^{\textit{i}\vec{K}\cdot\vec{r}} d^3r
\end{equation}
where $\vec{K}$ is a reciprocal lattice vector which is equivalent to a momentum transfer vector of scattering experiments.

We used the spin polarized linear augmented plane wave (LAPW) method \cite{num14} to obtain the electronic structure and spin density. The Fourier transformation of the plane wave expanded spin density in the interstitial region was added to the form factor for the angular decomposed spin density inside the muffin-tin spheres. To calculate the self-consistent spin density, we used a full-potential LAPW method \cite{num15} with a local density functional.\cite{num16} The quantity $R_{MT}\cdot K_{max}$, where $R_{MT}$ is the smallest muffin-tin radius and $K_{max}$ is the plane-wave cutoff, was set to 8.0 and 500 k-points in the irreducible Brillouin zone were generated for integration using the tetrahedron method.\cite{tetra} We used muffin-tin radii of 2.3, 2.1 and 2.1 \textit{a.u.} for the respective Sr, Fe and As spheres and 0.01 mRy/cell of the total energy as the convergence criterion. The structural parameters were taken from reported X-ray powder diffraction results.\cite{num17}
Since it is well known that calculated magnetic properties are very sensitive to the As positon \cite{num7,new1,new2}, we performed the magnetic form factor calculation with both the experimental and the energy minimized ``optimized'' As position. To obtain the optimized As position, As atoms were relaxed along \textit{z}-direction until the total force on As atom was less than 0.1 mRy/\textit{a.u.} .

To obtain the form factor of iron in SrFe$_2$As$_2$ we collected the integrated intensities of the rocking curves of nuclear and magnetic Bragg reflections to the highest order possible with the HB-1A spectrometer. The nuclear reflections, necessary to obtain the magnetic moment, were chosen to be as close as possible in the scattering angle $2\theta$ to that of the corresponding magnetic Bragg reflections. This reduces any errors due to geometry corrections, Debye-Waller factor (DWF), absorption effects, etc. The integrated intensity of a rocking curve of a nuclear Bragg reflection at a reciprocal lattice vector \textbf{q} in a crystal is given by:
\begin{eqnarray}
I_N &= \frac{V}{v_0^2}\Phi_0(\theta)|F_\texttt{N}(\textbf{q})|^2\frac{\lambda^3}{2\mu\sin 2\theta}\mbox{ e}^{-2W} \nonumber\\
& = \frac{C(q)|F(\textbf{q})|^2}{\sin 2\theta}
\end{eqnarray}
where $V$ is the scattering volume of the crystal, $v_0$ is the unit-cell volume, $\Phi_0(\theta)$ is the beam flux at the angle $\theta$, $F_\texttt{N}(\textbf{q})$ is the structure factor, $\lambda$ is the beam wavelength, $\mu$ is the absorption length, $\sin(2\theta)$ is the Lorentz factor for a rotating crystal, and $2W = \textbf{q}^2\langle u_\texttt{Q}\rangle^2$ is the DWF. We collect all the constants and unknown \textbf{q}-dependent factors in $C(\textbf{q})$, including the DWF. The integrated intensities of the rocking curves of the chosen nuclear Bragg reflections were normalized to the corresponding structure factors and the Lorentz factors. The data were also corrected for the fact that the rocking curves of nuclear Bragg peaks in our measurements include the twinned (\emph{h} 0 0) and (0 \emph{k} 0) orthorhombic domains, for normalization of the magnetic reflections which are due to the (\emph{h} 0 0) domain only. 
We assume $C(\textbf{q})$ is a DWF-like function (namely, a gaussian) and fit the data by the nonlinear least square technique to obtain a smooth $C(\textbf{q})$ function.
The refined parameter $\langle u_\texttt{Q}\rangle$ in this procedure is in good agreement with the corresponding values obtained in a powder diffraction study of SrFe$_2$As$_2$.\cite{num13}
\begin{table}[!ht]
\caption{The \textbf{q} values, integrated intensities, and the corresponding magnetic structure factors at 10 K of the magnetic reflections observed in single-crystal SrFe$_2$As$_2$ with the AF structure as shown Fig.\ \ref{struct1}.}
\label{Table1}
\begin{ruledtabular}
\begin{tabular} {ccccc}
 Reflection & \textbf{q} (\AA$^{-1}$)& Intensity & $|F_\texttt{M}|^2\sin^2\alpha$ & Form Factor\\
  &&&\\
\hline
(101)& 1.244 & 128.6$\pm$3 & 10.92    & 0.9170$\pm$0.0096  \\
(103)& 1.9084  & 242.1$\pm$4 & 41.53  & 0.8576$\pm$0.0170\\
(105)&  2.798  & 110.9$\pm$3 & 53.6   & 0.7085$\pm$0.0328 \\
(107)& 3.756   & 30.5$\pm$1  & 58.2   & 0.4954$\pm$0.0632 \\
(109)& 4.7417  & 7.91$\pm$0.58& 60.38 & 0.3204$\pm$0.1058\\
(301)& 3.441   & 1.19 $\pm$0.38 & 1.4 & 0.5678$\pm$0.0512\\
(303)& 3.733   & 6.57$\pm$0.39& 10.9  & 0.5273$\pm$0.0623\\
(305)& 4.257   & 5.28$\pm$0.28 &23.3  & 0.3799$\pm$0.0860\\
\end{tabular}
\end{ruledtabular}
\end{table}

Similarly, the scattered intensity of a magnetic Bragg reflection can be expressed as:
\begin{equation}
I_M = C(q) (\gamma r_0\frac{1}{2}gS)^2f(q)^2|F_\texttt{M}(\textbf{q})|^2\sin^2\alpha\frac{1}{\sin 2\theta}
\label{Eq-magnetic}
\end{equation}
where $f(q)$ is the magnetic form factor at the magnetic reciprocal lattice \textbf{q}, $F_\texttt{M}(\textbf{q})$ is the magnetic structure factor\cite{num18}, and $\sin^2\alpha=1-(\hat{\bf{q}}\cdot \hat{\bf{\mu}})^2$ where $\hat{\bf{q}}$ and $\hat{\bf{\mu}}$ are the unit vectors along the scattering vector (reciprocal vector) and the direction of the moment, respectively (in this case $\hat{\bf{\mu}}=$ (1 0 0)).
In Table\ \ref{Table1}, we list all the observed magnetic reflections, their integrated intensities, and the corresponding magnetic structure factors taking the magnetic structure as shown in Fig.\ \ref{struct1}. Using these parameters in Table\ \ref{Table1} and Eq.\ (\ref{Eq-magnetic}) we calculated the value of the form factor $gSf(|\textbf{q}|)$ as shown below in Fig.\ \ref{formfactor1}(b) (blue triangles). 

To estimate the magnetic moment we fit the measured form factor with Fe$^{2+}$ ionic form factor to extrapolate the data to \textbf{q}=0 yielding \textit{M}= 1.07(2) $\mu_\texttt{B}$. We also used the shape of the calculated form factor (described below) to extrapolate the data yielding a magnetic moment of \textit{M}= 1.04(1) $\mu_\texttt{B}$.

We now turn to the theoretical calculations of the spin density and magnetic form factor. The theoretical energy minimization process of relaxing the As position resulted in $z_{As}= 0.3499$ while the starting experimental position was $z_{As}=0.3612$. The distance between Fe-As was decreased about 3\% (from 2.392 \AA  to 2.315 \AA) and the magnetic moment on the Fe atoms was changed from 1.676 $\mu_\texttt{B}$ to 0.879 $\mu_\texttt{B}$ (at the optimized theoretical minimum energy \textit{z}$_{As}$ position). Table\ \ref{Table2} shows the Fe \textit{d}-orbital decomposed charges and moments. Although there are variations among the occupied electronic orbitals, magnetic moments are fairly well distributed across all of the orbitals. This would not be expected if one or more of the orbitals were ``localized''. Indeed, all of orbitals contribute to a broad density of states.

\begin{table}[htbp]
	\centering
\caption{Fe \textit{d}-orbital electron occupation and magnetic moment orbital decomposition. }
\label{Table2}

\begin{tabular}{|c||c|c||c|c||c|c|}\hline 
&\multicolumn{2}{c||}{spin up} &\multicolumn{2}{c||}{spin dn} &\multicolumn{2}{c|}{Magnetic Moment}\\  
& \multicolumn{2}{c||}{(electrons) }&\multicolumn{2}{c||}{(electrons) }& \multicolumn{2}{c|}{($\mu_\texttt{B}$)} \\ \hline
&exp z$_{As}$ &opt z$_{As}$ &exp z$_{As}$ &opt z$_{As}$ &exp z$_{As}$ &opt z$_{As}$ \\ \hline
d      & 3.810  & 3.450 & 2.142 & 2.574 & 1.668 & 0.876 \\  \hline
d$_{{z}^{2}}$ & 0.719  & 0.618 & 0.495 & 0.513 & 0.224 & 0.105 \\ \hline
d$_{{x}^{2}}$$_{-}$$_{{y}^{2}}$ & 0.757 &0.684 &0.402 &0.491 &0.355 &0.193 \\  \hline
d$_{xy}$ & 0.769 &0.700 & 0.502 & 0.568 & 0.267 & 0.132 \\  \hline
d$_{xz}$ & 0.767 &0.718 & 0.354 & 0.476 & 0.413 & 0.242 \\  \hline
d$_{yz}$ & 0.798 &0.730 & 0.389 & 0.526 & 0.409 & 0.204 \\  \hline 
\end{tabular}
\end{table}

\begin{figure}[htbp]
	\centering
   \includegraphics[width=0.50\textwidth]{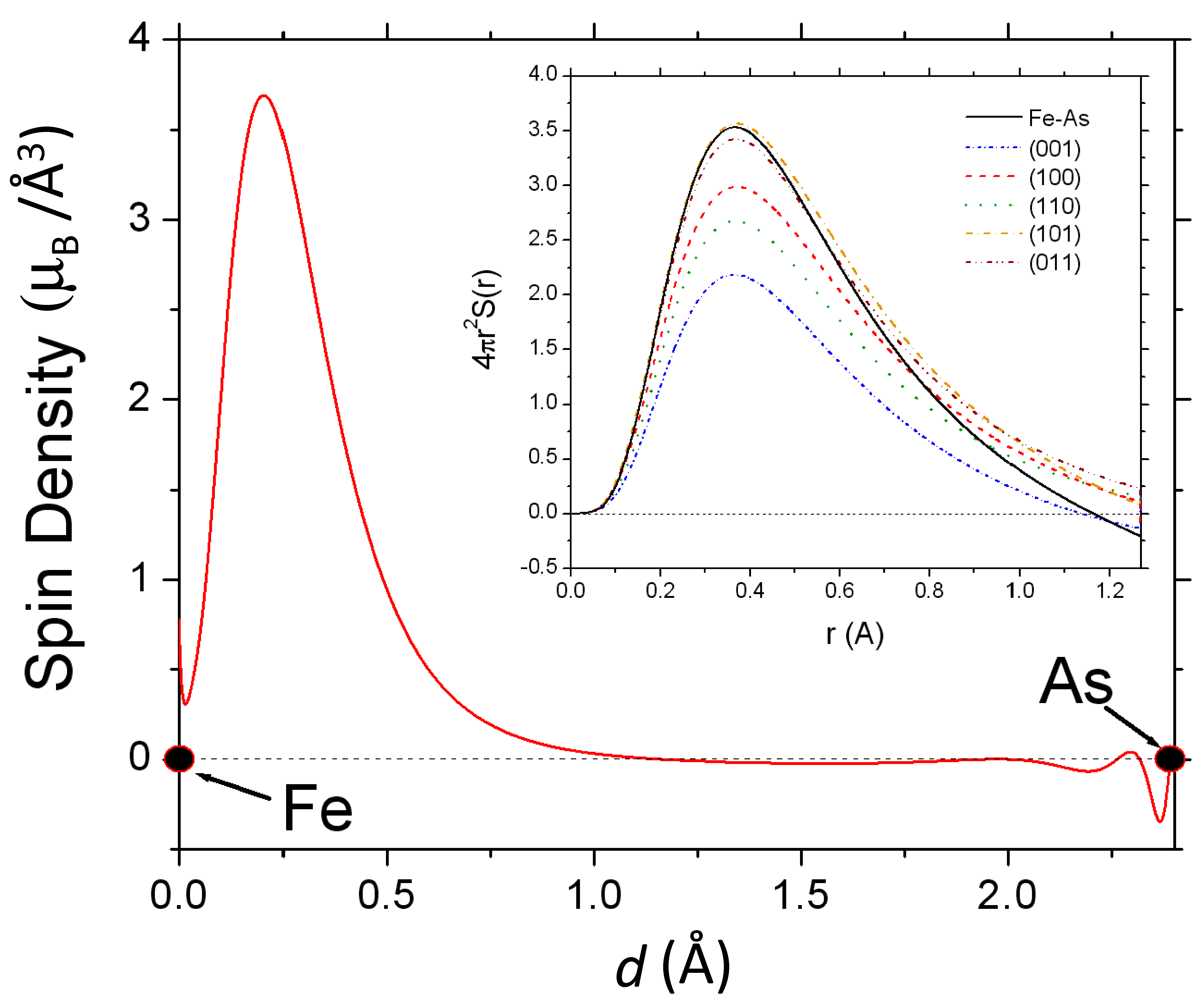}
	\caption{(Color online) Spin density along the Fe-As direction as calculated from LDA with the experimental atomic position. Although As does not have a net magnetic moment in the antiferromagnetic structure, its spin density distribution depends on the magnetic ordering of Fe. The inset shows spin densities inside the Fe muffin-tin sphere along specific directions demonstrating substantial anisotropy.}
	\label{spin1d1}
\end{figure}

Figure\ \ref{spin1d1} shows the spin density along Fe to As. It indicates that although the spin density is strongly concentrated inside Fe atoms, the spin density inside the As muffin-tin sphere is affected by the surrounding Fe atoms through hybridization. With antiferromagnetic ordering (see Fig.\ \ref{struct1}), the As atoms do not have a net moment. There are, however, spin distributions surrounding the As site that depend on the surrounding Fe atoms. The inset of Fig.\ \ref{spin1d1} shows the radially weighted spin density inside the  Fe muffin-tin sphere with a high density along the Fe-As direction and a significant drop in spin density along the \textit{z}-axis. It clearly shows anisotropy of the spin density inside of the Fe muffin-tin sphere.

\begin{figure}[htbp]
	\centering
   \includegraphics[width=0.50\textwidth]{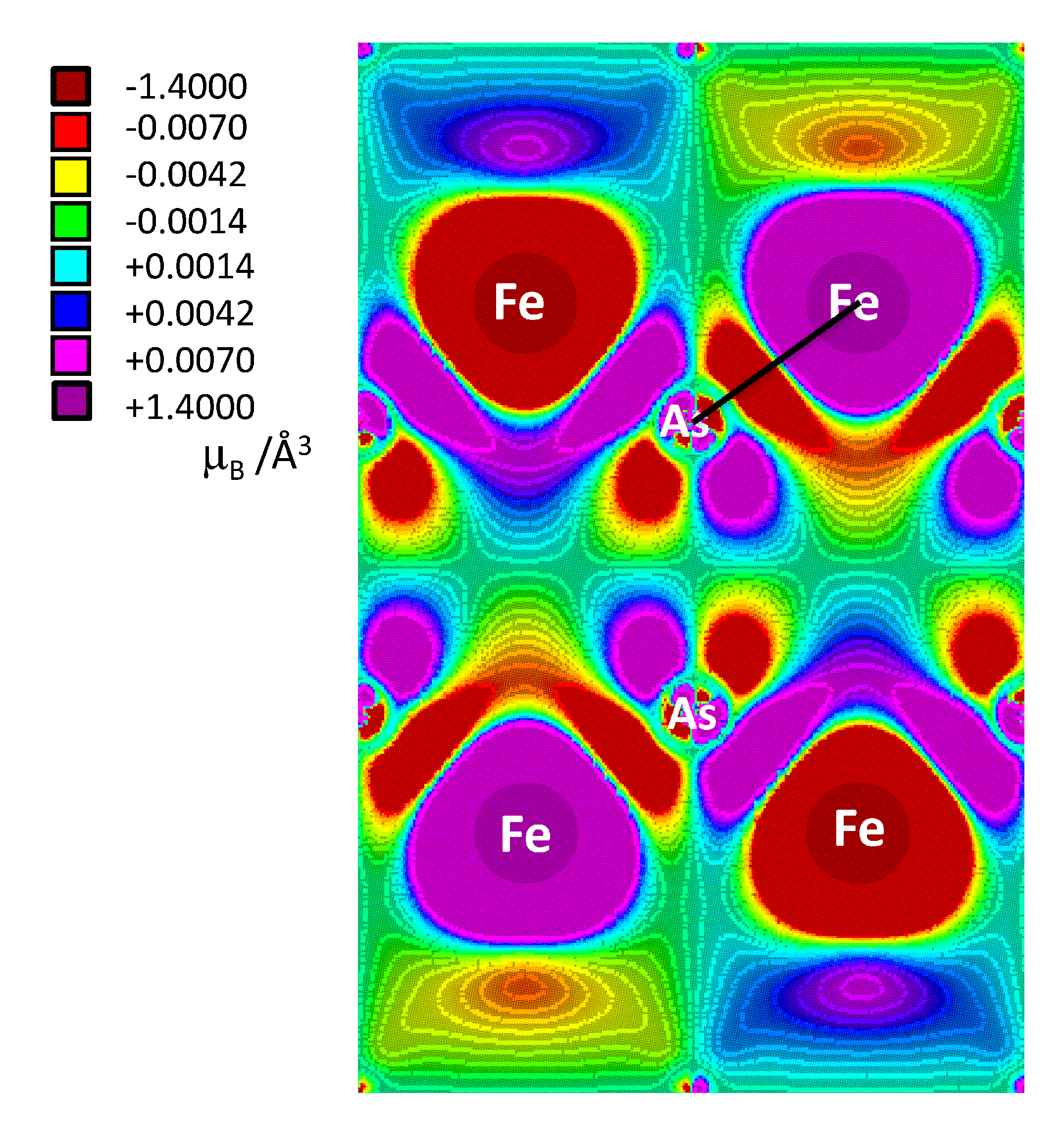}
	\caption{(Color online) 2-dimensional spin density plot indicating hybridization between Fe and As atoms and anisotropy. Since the spin density is strongly concentrated in the small radius inside Fe atoms(See Fig.\ \ref{spin1d1}) a small spin density contour range was chosen to reveal the low spin density structure. The black line between Fe and As is the axis for Fig. \ref{spin1d1}}
	\label{spin2d1}
\end{figure}

Figure\ \ref{spin2d1} shows spin density contours in the gray plane shown in Fig.\ \ref{struct1}. Since we are interested in the hybridization between Fe and As, we have chosen a plane that is determined by (1,1,0) and (0,0,1) vectors. It shows the antiferromagnetically ordered Fe atoms and the effect of the ordered Fe atoms on As atoms. The hybridization between Fe \textit{d} electrons and As \textit{p} electrons produces an anisotropic spin density distribution around the As atoms.  

\begin{figure}[htbp]

	\centering		
\includegraphics[width=0.50\textwidth]{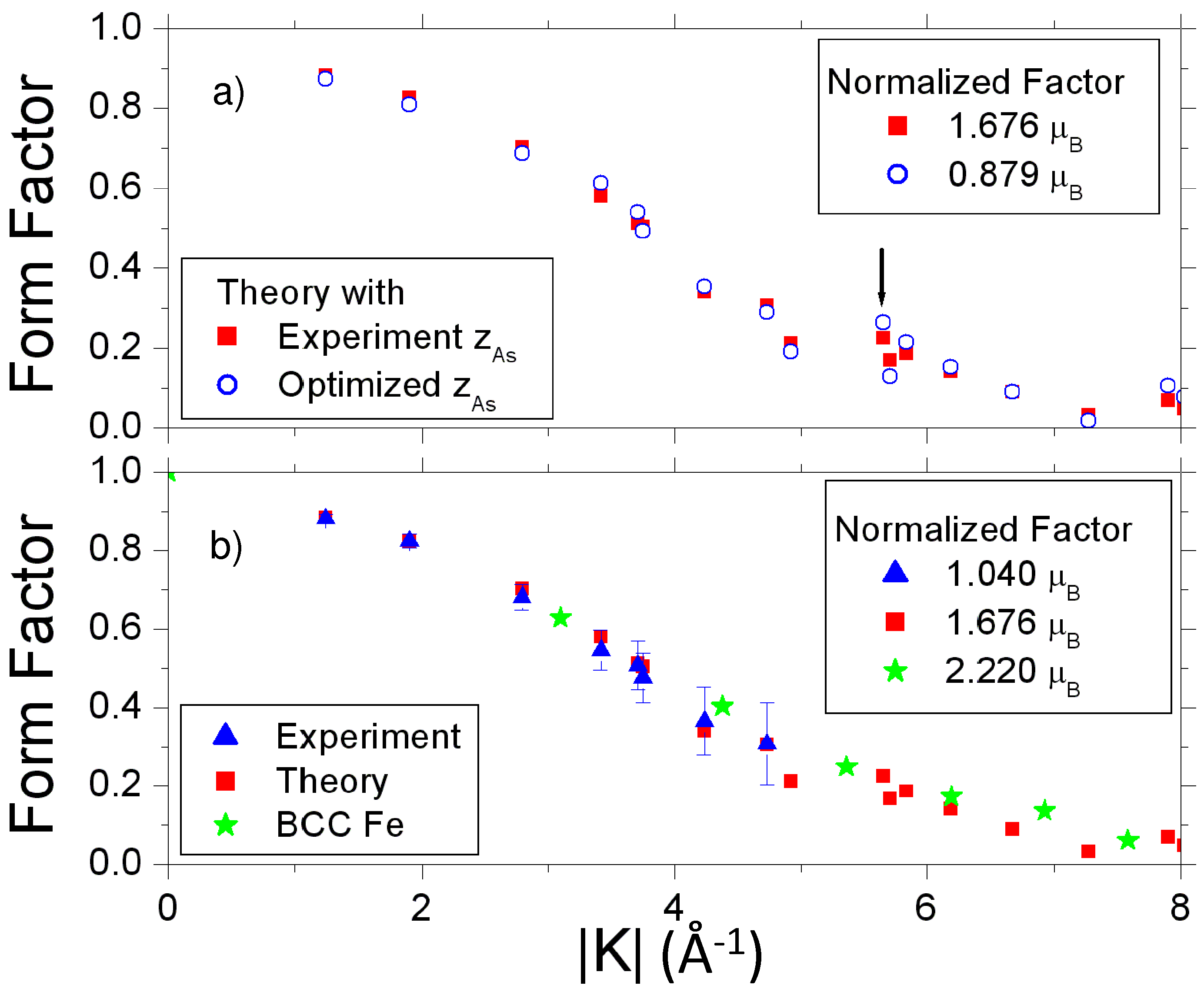}
	\caption{(Color online) a). Comparison of the form factors calculated with the experimental atomic positions (\textit{z}$_{As}$ = 0.3612) and theoretical energy optimized one (\textit{z}$_{As}$ = 0.3499). The arrow points to scattering vectors that provide evidence for fairly strong anisotropy in the form factor. b). Experimental form factors for SrFe$_{2}$As$_{2}$ and BCC Fe, and the calculated form factors with the experimental As position. }
	\label{formfactor1}
\end{figure}

Figure\ \ref{formfactor1} a) shows the form factors that were calculated with the experimental As position (red filled squares) and the optimized As position (blue open circles). Data were normalized with the form factor at $\left|K \right|$=0 corresponding to the magnetic moment of the Fe atom. The arrow points out the two scattering vectors with similar magnitude but different intensities - demonstrating an anisotropy increase for the ``optimized'' As position with a smaller magnetic moment. The K vector which has higher intensity is (5,0,1) and the lower intensity one is (3,0,9). Figure\ \ref{formfactor1} b) includes a comparison between the experimental form factor of SrFe$_{2}$As$_{2}$, that of bcc iron\cite{num17}, and the calculated one using the experimental As position. Even though SrFe$_{2}$As$_{2}$ has a much more complicated electronic structure than bcc Fe, their normalized form factors are quite similar, with differences showing up at larger and more difficult to measure scattering vectors. The similarity with bcc Fe is particularly evident in the total 3\textit{d} charge density within the muffin-tin sphere. This is 5.95 electrons for the experimental atomic positions in SrFe$_{2}$As$_{2}$, 6.02 electrons for the corresponding calculated energy optimized As position and 5.83 electrons for bcc Fe calculated with the same muffin-tin radius.   

We have determined the magnetic form factor of Fe in SrFe$_{2}$As$_{2}$ by neutron diffraction experiments and deduced the Fe magnetic moment as 1.04(1)$\mu_\texttt{B}$. We also calculated the magnetic form factor by first principles electronic structure methods using both the experimental and optimized As positions. While the magnitude of the calculated magnetic moments strongly depends on the As position, the normalized magnetic form factors were remarkably similar with each other and also agreed well with the normalized  magnetic form factors from the experiment. The largest difference between the two theoretical calculations was the noticeable increase in the spin anisotropy for the optimized As position. The experiments were limited by spectrometer geometry and did not measure far enough in reciprocal space to access these contributions. 

Research supported by the U.S. Department of Energy, Office of Basic Energy Sciences, Division of Materials Sciences and Engineering under Contract No. DE-AC02-07CH11358.

\end{document}